\documentclass[10pt, conference]{IEEEtran}
\IEEEoverridecommandlockouts

\usepackage[utf8]{inputenc}
\usepackage{amsmath,amssymb,amsfonts}
\usepackage{algorithmic}
\usepackage{graphicx}
\usepackage{textcomp}
\usepackage{xcolor}
\usepackage{breqn}
\usepackage{siunitx}
\usepackage[numbers,sort]{natbib}

\usepackage{booktabs}
\usepackage{multirow}

\makeatletter
\let\MYcaption\@makecaption
\makeatother

\usepackage{subcaption}
\makeatletter
\let\@makecaption\MYcaption
\makeatother

\begin{document}

\title{End-to-End Integration of Speech Emotion Recognition with Voice Activity Detection using Self-Supervised Learning Features}

\author{\IEEEauthorblockN{Natsuo Yamashita}
\IEEEauthorblockA{\textit{Research \& Development Group,} \\
\textit{Hitachi, Ltd.}\\
Tokyo, Japan \\
natsuo.yamashita.gh@hitachi.com}
\and
\IEEEauthorblockN{Masaaki Yamamoto}
\IEEEauthorblockA{\textit{Research \& Development Group,} \\
\textit{Hitachi, Ltd.}\\
Tokyo, Japan \\
masaaki.yamamoto.af@hitachi.com}
\and
\IEEEauthorblockN{Yohei Kawaguchi}
\IEEEauthorblockA{\textit{Research \& Development Group,} \\
\textit{Hitachi, Ltd.}\\
Tokyo, Japan \\
yohei.kawaguchi.xk@hitachi.com}
}

\maketitle

\begin{abstract}

Speech Emotion Recognition (SER) often operates on speech segments detected by a Voice Activity Detection (VAD) model. However, VAD models may output flawed speech segments, especially in noisy environments, resulting in degraded performance of subsequent SER models. To address this issue, we propose an end-to-end (E2E) method that integrates VAD and SER using Self-Supervised Learning (SSL) features. The VAD module first receives the SSL features as input, and the segmented SSL features are then fed into the SER module. Both the VAD and SER modules are jointly trained to optimize SER performance. Experimental results on the IEMOCAP dataset demonstrate that our proposed method improves SER performance. Furthermore, to investigate the effect of our proposed method on the VAD and SSL modules, we present an analysis of the VAD outputs and the weights of each layer of the SSL encoder.

\end{abstract}


\section{Introduction}
\label{sec:intro}
Speech Emotion Recognition (SER) is the task of identifying and classifying emotional states expressed in spoken language \cite{review}. 
It is an essential field within the expansive domain of affective computing \cite{affective_computing} and human-computer interaction \cite{HI}, with a variety of real-world applications, including healthcare \cite{healthcare}, customer service \cite{customer_service}, and marketing \cite{marketing}. 

Typical SER methods \cite{ser_conventional_1,ser_conventional_2,ser_conventional_3,ser_conventional_4} begin by extracting low-level descriptive features, such as prosodic characteristics and spectral features from speech signals, which are then fed into machine learning models. 
With recent advances in deep learning, especially Transformer framework \cite{transformer}, a considerable number of studies \cite{ssl_based_1,ssl_based_2,ssl_based_3} have focused on utilizing pre-trained self-supervised learning (SSL) models such as wav2vec 2.0 \cite{wav2vec2}, HuBERT \cite{hubert}, and WavLM \cite{wavlm} as feature extractors. 
The SER models using these SSL features (SSL-SER) have shown greatly improved performance in SER and other downstream tasks through fine-tuning \cite{finetuning_1,finetuning_2,finetuning_3,superb}. 

In practical applications of SER, it is common practice to employ a Voice Activity Detection (VAD) model, which detects the speech segments in a given audio sequence, before feeding the audio into a SER model \cite{vad_review}. 
This pre-processing step aims to identify when specific emotions are expressed, mitigate the effect of noise on the SER model, and reduce the computational costs for post-processing.
Though deep learning-based VAD models \cite{marblenet,vad_deeplearning_1,vad_deeplearning_2} have improved performance compared to traditional schemes based on statistics of speech \cite{vad_conventional_1,vad_conventional_2,webrtcvad}, they can still generate flawed output, especially in noisy environments.
This output may be fragmented, miss emotional features, or contain noise at the beginning, in the intervals, or at the end, resulting in degraded SER performance.
In the context of Automatic Speech Recognition (ASR), this is a less crucial issue, because ASR, which is designed to extract meaningful linguistic information, can often compensate for minor segmentation errors.
In contrast, SER is more susceptible to flawed VAD outputs because it relies on detecting subtle variations and nuances in speech, such as tone, pitch, rhythm, intensity, and speed \cite{ser_noise}. 
Since VAD models are trained to distinguish speech from non-speech but are not optimized for SER performance, they sometimes fail to generate optimal output for the subsequent SER models.
Even if speech segments are correctly identified, some parts may contain rich emotional parts while others may not, and the latter can lead to a decline in the accuracy of SER. 

In this paper, to address the problem of degraded SER performance due to flawed VAD outputs in noisy environments, we propose a method that integrates VAD and SER modules using SSL features in an end-to-end (E2E) manner.
SSL features are first input into the VAD module, and then the segmented SSL features are fed into the SER module. 
Both the VAD and SER modules are jointly trained to optimize SER loss.
This approach allows the VAD module to be trained to include more emotional speech segments that are important for SER, while the SER module is trained to be robust against flawed segments from the VAD module. 
As our experimental results demonstrate, the proposed method improves SER performance in noisy environments on the IEMOCAP dataset. 


\section{Proposed approach}
\label{sec:proposal}
\begin{figure}[t]
\begin{minipage}[b]{1.0\linewidth}
  \centering
  \centerline{\includegraphics[width=0.9\columnwidth]{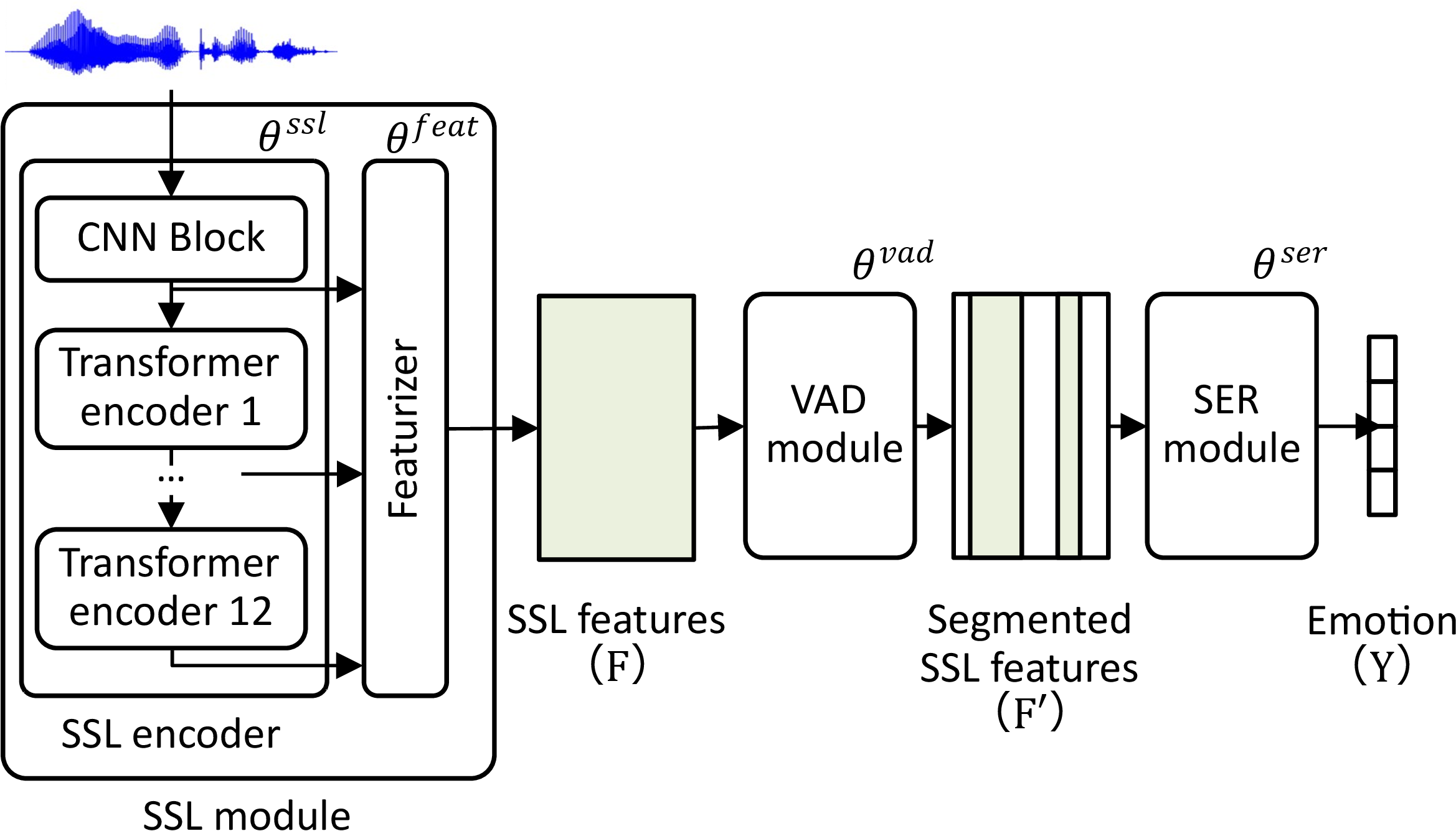}}
\end{minipage}
\caption{Overview of the proposed end-to-end approach composed of SSL, VAD, and SER modules.}
\label{fig:proposal}
\end{figure}

The overall architecture of the proposed approach is shown in Fig.~\ref{fig:proposal}, which consists of the SSL, VAD, and SER modules.
It is worth noting that the choice of network architecture for each module is not restricted to any specific one.
\subsection{SSL module}
\label{ssec:ssl}

In this study, we employ SSL models as common feature extractors for both the following VAD and SER modules, due to their well-known generalizability and accessibility across various speech processing tasks \cite{superb}. 
We denote the feature extraction process of the SSL module as follows:
\begin{align}
    \mathbf{F} &= \text{SSL}(\mathbf{X}; \,\theta^{\text{ssl}}, \,\theta^{\text{feat}}),
\end{align}
where $\mathbf{X}$ is an input utterance, $\mathbf{F}$ is the SSL features, and $\theta^{\text{ssl}}$ and $\theta^{\text{feat}}$ represent the parameters of the SSL encoder and the Featurizer, respectively.
Given an input waveform, the SSL encoder, consisting of a Convolutional Neural Network (CNN) block and 12 Transformer encoder blocks, extracts a frame sequence of 768-dimensional speech features with a frame shift of \SI{20}{\milli\second}.

Research \cite{wavlm,superb,ssl_layer} has shown that intermediate representations of such foundation models contain information useful for different tasks. 
Therefore, the Featurizer computes the weighted-sum of embeddings from the 13 hidden states of the SSL encoder.

\subsection{VAD module}
\label{ssec:vad}
While conventional VAD methods \cite{marblenet,vad_deeplearning_1} often use spectrogram features such as log Mel-Filterbanks (Fbank) and Mel-Frequency Cepstral Coefficients (MFCC), very recent work \cite{ssl_vad} has shown that a VAD architecture based on wav2vec 2.0 outperforms previous works \cite{vad_deeplearning_1,silerovad}. 
In this study, we employ the VAD module using SSL features (SSL-VAD) and investigate the use of not only wav2vec 2.0 but also HuBERT and WavLM for our end-to-end approach.
If we denote the VAD outputs indicating speech/non-speech as $\mathbf{S}$, the SSL features as $\mathbf{F}$, and the segmented SSL features as $\mathbf{F'}$, we can write the process of the VAD module:
\begin{align}
    \mathbf{S} &= \text{VAD}(\mathbf{F}; \,\theta^{\text{vad}}), \\
    \mathbf{F'} &= \mathbf{F}\odot\mathbf{S}
\end{align}
where $\theta^{\text{vad}}$ represents the parameters of the VAD module and $\odot$ is the Hadamard product operator.

The VAD module consists of four 1D convolution layers with a hidden dimension of 256 and leaky ReLU activation, and a Fully-Connected (FC) layer with softmax activation. 
It detects frame-level speech/non-speech, assigning a label of 1 or 0 to each frame using an argmax operation.
The segmented SSL features are then calculated by the Hadamard product between the SSL features and the broadcast outputs of the VAD process.

\subsection{SER module}
\label{ssec:ser}
We use the segmented SSL features from the VAD module as input for the SER module.
The SER process can be written as:
\begin{align}
    \hat{\mathbf{Y}} &= \text{SER}(\mathbf{F'}; \,\theta^{\text{ser}}),
\end{align}
where $\hat{\mathbf{Y}}$ is a predicted emotion label and $\theta^{\text{ser}}$ represents the parameters of the SER module. 
Given the segmented 768-dimensional SSL features, the SER module first applies dimensionality reduction from 768 to 256, followed by average pooling. 
The representations are then processed through three 1D convolution layers and a FC layer with ReLU activation. 
A subsequent self-attention pooling layer \cite{self_attention_pooling} aggregates features along the time axis, which are then fed into a FC layer with ReLU activation and a FC layer with softmax activation for emotion classification.


\subsection{End-to-end training}
\label{ssec:e2e}
Some studies attempted to combine VAD and SER modules in a cascade manner \cite{vad_ser} or using a multi-task approach \cite{multitask}, but end-to-end training of VAD and SER modules (E2E SSL-VAD-SER) has not been fully investigated.
In this study, we propose an end-to-end approach that jointly optimizes the VAD and SER modules for SER loss using the SSL features, to address the issue of degradation in SER performance due to flawed VAD outputs in noisy environments.
Our entire end-to-end approach can be described as follows:
\begin{dmath}
    \hat{\mathbf{Y}} = \text{SER}(\,\text{SSL}(\text{X}; \,\theta^{\text{ssl}}, \,\theta^{\text{feat}})\odot\text{VAD}(\,\text{SSL}(\text{X}; \,\theta^{\text{ssl}}, \,\theta^{\text{feat}}); \, \theta^{\text{vad}}); \,\theta^{\text{ser}}),
\end{dmath}

To give effective feedback between the VAD and SER modules from the start of end-to-end training, we initialize each parameter with $\hat{\theta}^{\text{ssl}}$ from publicly available pre-trained models and $\hat{\theta}^{\text{feat}}$, $\hat{\theta}^{\text{vad}}$, and $\hat{\theta}^{\text{ser}}$ obtained by pre-training the SSL-VAD and SSL-SER, respectively. 
Then, $\hat{\theta}^{\text{feat}}$, $\hat{\theta}^{\text{vad}}$, and $\hat{\theta}^{\text{ser}}$ are fine-tuned for SER loss to achieve better performance, while $\hat{\theta}^{\text{ssl}}$ is frozen in this study to reduce computational costs.
Note that this end-to-end approach allows for fine-tuning modules even on speech data that has only emotion labels without speech/non-speech annotations for VAD.

\section{Experimental setup}
\label{sec:exp}
\subsection{Dataset}
\label{ssec:data}
We used the IEMOCAP dataset \cite{iemocap} which has approximately 12 hours of English speech including 5 dyadic conversational sessions between two actors.
There are in total 151 dialogues, including 10,039 utterances. 
The IEMOCAP provides the timestamps for each utterance in a dialogue, as well as word-level alignments for each utterance.
The utterances contain silence at the beginning, between words, and at the end. 
The VAD module is trained based on these speech/non-speech alignments, with $\SI{40}{\percent}$ of the segments being non-speech.
For SER, we merged emotion class ``excited'' with ``happy'' and used audio annotated with one of four labels, which are happy, sad, neutral, and angry. 

During evaluation, to simulate a real-world scenario, we extended the original utterances by concatenating the intervals before and after them.
Specifically, we used the timestamps from the end of the previous utterance to the start of the next utterance, ensuring there was no overlap by subtracting $\SI{0.1}{\second}$.
These extended utterances averaged $\SI{11.7}{\second}$, while the original utterances averaged $\SI{4.5}{\second}$.
The original and extended utterances were contaminated with noise, randomly selected from the 37 recordings annotated as background noise in the MUSAN corpus \cite{musan} at Signal-to-Noise Ratio (SNR) levels of $\{10, 5, 0, -5, -10\}$ \si{\dB}.
Speakers in the test set were excluded from the training and validation sets. 
Additionally, we made sure that utterances from the same dialogue were either all in the training set or all in the validation set.


\subsection{Model configuration}
\label{ssec:model}
We used pre-trained SSL models including wav2vec 2.0 \textsc{Base} \cite{wav2vec2}, HuBERT \textsc{Base} \cite{hubert}, and WavLM \textsc{Base+} \cite{wavlm}, which are publicly available. 
For simplicity, the term ``\textsc{Base}'' will be omitted hereafter.
%
Each model has approximately 95 million parameters.
Wav2vec 2.0 and HuBERT were trained with the concatenation of the \textit{train-clean-100}, \textit{train-clean-360}, and \textit{train-other-500} subsets from the LibriSpeech dataset \cite{librispeech}. WavLM+ was trained with the Libri-Light \cite{libri_light}, GigaSpeech \cite{gigaspeech}, and VoxPopuli datasets \cite{voxpopuli}. 
The SSL-VAD and SSL-SER were pre-trained and initialized using these SSL models.
They were trained using the Adam optimizer \cite{adam} with a fixed learning rate of $1\times10^{-4}$ and a batch size of 8, without data augmentation, for at most 100 epochs.
In the proposed approach, the modules were then fine-tuned using SER loss with the same optimizer algorithm and learning rate. 
During the training of each system, the parameters of the SSL encoder were not updated.

\section{Results}
\label{sec:results}
\subsection{Performance on SER}
In this section, we present the evaluation results on the IEMOCAP dataset.
We computed Unweighted Accuracy (UA) and Weighted Accuracy (WA) on the test sets, as shown in Table \ref{tab:ser_performance}. 
UA is the average recall across all categories and WA is the total number of correct predictions divided by the total number of samples.
First, we report the performance of the SSL-SER, which does not include the VAD module, as a reference (Condition 1). 
It was evaluated using the original utterances, as in previous studies \cite{superb,finetuning_1}.
For the evaluation of Conditions 2-7, we used the extended utterances containing intervals before and after, as described in section \ref{ssec:data}. 
Conditions 2 and 3 serve as baselines, while Conditions 4-7 represent our proposed approaches.
The performance of Condition 2, which was based on the same architecture as Condition 1 but was evaluated with the extended utterances, shows a dramatic degradation compared to Condition 1 due to the effect of the noisy extended utterances.
In Condition 3, we implemented MarbleNet \cite{marblenet}, one of the common deep learning-based VAD models, as a pre-processing VAD step for the SSL-SER. 
MarbleNet utilized MFCC with 64 dimensions of mel-filter bank, a \SI{25}{\milli\second} window size and a \SI{10}{\milli\second} overlap. 
MarbleNet and the SSL-SER were individually trained and the segmented speeches detected by MarbleNet were fed into the SSL-SER. 
The results of Condition 3 show that SER performance was degraded when MarbleNet was used, compared to Condition 2.
This indicates that simply combining a VAD model with the SSL-SER in noisy environments can cause detrimental effects on SER performance due to flawed VAD outputs.

\begin{table}[t]  
    \centering  
        \caption{SER performance (\si{\percent}UA, \si{\percent}WA) for the referential condition (Condition 1) using the original utterances (original utt.), and for the baselines (Conditions 2 and 3) and proposed fine-tuning (FT) approaches (Conditions 4-7) using the extended utterances.}
    \resizebox{0.98\linewidth}{!}{%
        \tabcolsep = 4pt
        \begin{tabular}{clcccccc}  
            \toprule  
            & & \multicolumn{2}{c}{wav2vec 2.0}&\multicolumn{2}{c}{HuBERT}&\multicolumn{2}{c}{WavLM+} \\  
            \cmidrule(l{1mm}r{1mm}){3-4} \cmidrule(l{1mm}r{1mm}){5-6} \cmidrule(l{1mm}r{1mm}){7-8} 
            ID & Condition
            & UA & WA & UA & WA & UA & WA \\  
            \midrule  
            \midrule  
            1 & SSL-SER (original utt.) & 58.0 & 59.5 & 58.7 & 60.2 & 63.3 & 64.7 \\  
            \midrule  
            2 & SSL-SER &  46.0 & 49.9 & 42.6 & 47.4 & 43.8 & 48.2 \\  
            3 & MarbleNet-SSL-SER & 40.3 & 39.6 & 40.2 & 39.4 & 42.8 & 42.1 \\  
            \midrule  
            4 & SSL-VAD-SER& 49.3 & 47.0 & 45.5 & 49.0 & 44.6 & 41.7 \\  
            5 & E2E SSL-FT.VAD-SER & 49.7 & 48.0 & 47.4 & 50.5 & 46.2 & 43.6 \\  
            6 & E2E SSL-VAD-FT.SER & 50.2 & 53.1 & 48.9 & 52.1 & 49.0 & 53.0 \\  
            7 & E2E SSL-FT.VAD-FT.SER & \textbf{50.6} & \textbf{53.7} & \textbf{51.6} & \textbf{54.2} & \textbf{51.4} & \textbf{54.9} \\  
            \bottomrule  
        \end{tabular}%
    }
    \label{tab:ser_performance}  
\end{table}  


In Conditions 4-7, we explored four different fine-tuning strategies for our approach. 
Condition 4 consists of the individually pre-trained SSL-VAD and SSL-SER without fine-tuning. 
The results of Conditions 3 and 4 show that replacing the MFCC-based MarbleNet with the SSL-VAD improved SER performance, especially when using wav2vec 2.0 and HuBERT.
In Conditions 5-7, all trainable parameters were directly optimized for SER loss in an end-to-end manner. 
In Condition 5, $\hat{\theta}^{\text{feat}}$ and $\hat{\theta}^{\text{vad}}$ were fine-tuned while the other parameters were frozen.
The performance of Condition 5, on all the SSL models, shows improvement compared to Condition 4.
This suggests that the VAD module optimized only for speech/non-speech detection is not always optimal for SER, whereas our approach successfully improved the VAD outputs for SER.
In Condition 6, $\hat{\theta}^{\text{feat}}$ and $\hat{\theta}^{\text{ser}}$ were fine-tuned, while the other parameters were frozen.
Condition 6, with any SSL model, also improved SER performance compared to Condition 4, indicating the importance of fine-tuning on the segmented SSL features rather than only on the original utterances.
In Condition 7, $\hat{\theta}^{\text{feat}}$, $\hat{\theta}^{\text{vad}}$, and $\hat{\theta}^{\text{ser}}$ were fine-tuned and $\hat{\theta}^{\text{ssl}}$ was frozen. 
We observe that Condition 7 further improved SER performance on all the SSL models, indicating that our end-to-end approach successfully jointly fine-tuned both the VAD and SER modules for SER.
For example, comparing Condition 3 and 7 on WavLM+, the UA increased from \SI{42.8}{\percent} to \SI{51.4}{\percent} and the WA increased from \SI{42.1}{\percent} to \SI{54.9}{\percent}.

\subsection{Analysis of VAD outputs}

\begin{table}[t]  
    \centering  
    \caption{VAD performance with Accuracy (\si{\percent}Acc), Precision (\si{\percent}Prec), and Recall (\si{\percent}Rec) of MarbleNet-SSL-SER (Condition 3), SSL-VAD-SER (Condition 4), and E2E SSL-FT.VAD-FT.SER (Condition 7) on HuBERT and WavLM+.}  
    \resizebox{0.9\linewidth}{!}{%
        \begin{tabular}{llcccc}  
            \toprule  
            SSL Model & Condition & Acc & Prec & Rec \\  
            \midrule  \midrule  
             & MarbleNet-SSL-SER & 87.5 & 77.2 & 88.5 \\  
            \midrule  
            \multirow{2}{*}{HuBERT} &  SSL-VAD-SER & 90.0 & 74.9 & 92.2 \\  
            & E2E SSL-FT.VAD-FT.SER & 90.0 & 75.6 & 91.8  \\  
            \midrule  
            \multirow{2}{*}{WavLM+} & SSL-VAD-SER & 90.2 & 74.3 & 92.6  \\  
            & E2E SSL-FT.VAD-FT.SER & 89.6 & 73.8 & 93.0 \\  
            \bottomrule  
        \end{tabular}%
    }
    \label{tab:vad}
\end{table}

\begin{figure}[t]
    \centering
    \begin{minipage}[b]{0.43\columnwidth}
    
        \centering
        \includegraphics[width=0.85\columnwidth]{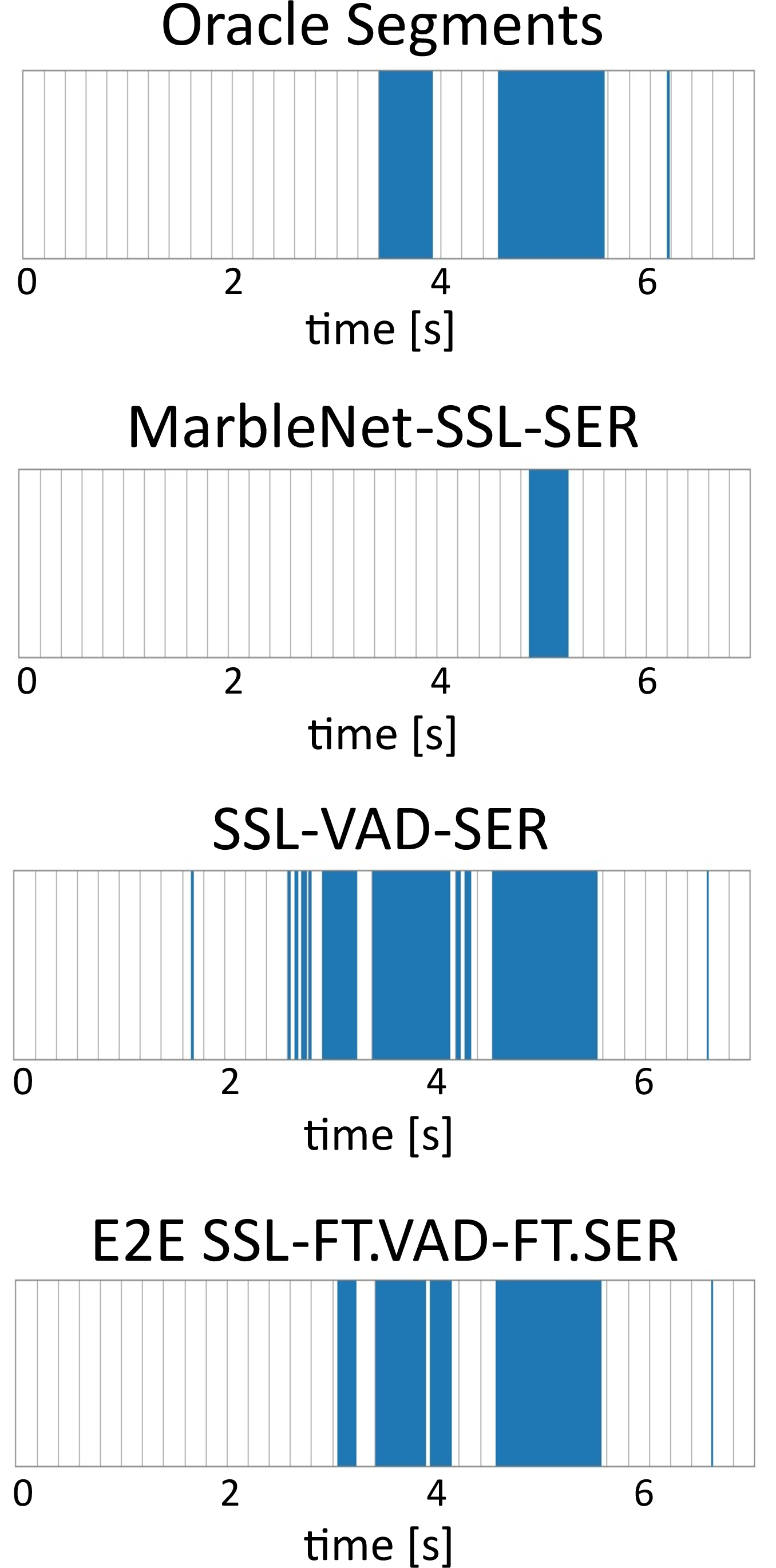}
        \subcaption{HuBERT}
        \label{fig:vad_hubert}
    \end{minipage}
    \begin{minipage}[b]{0.535\columnwidth}
        \centering
        \includegraphics[width=0.85\columnwidth]{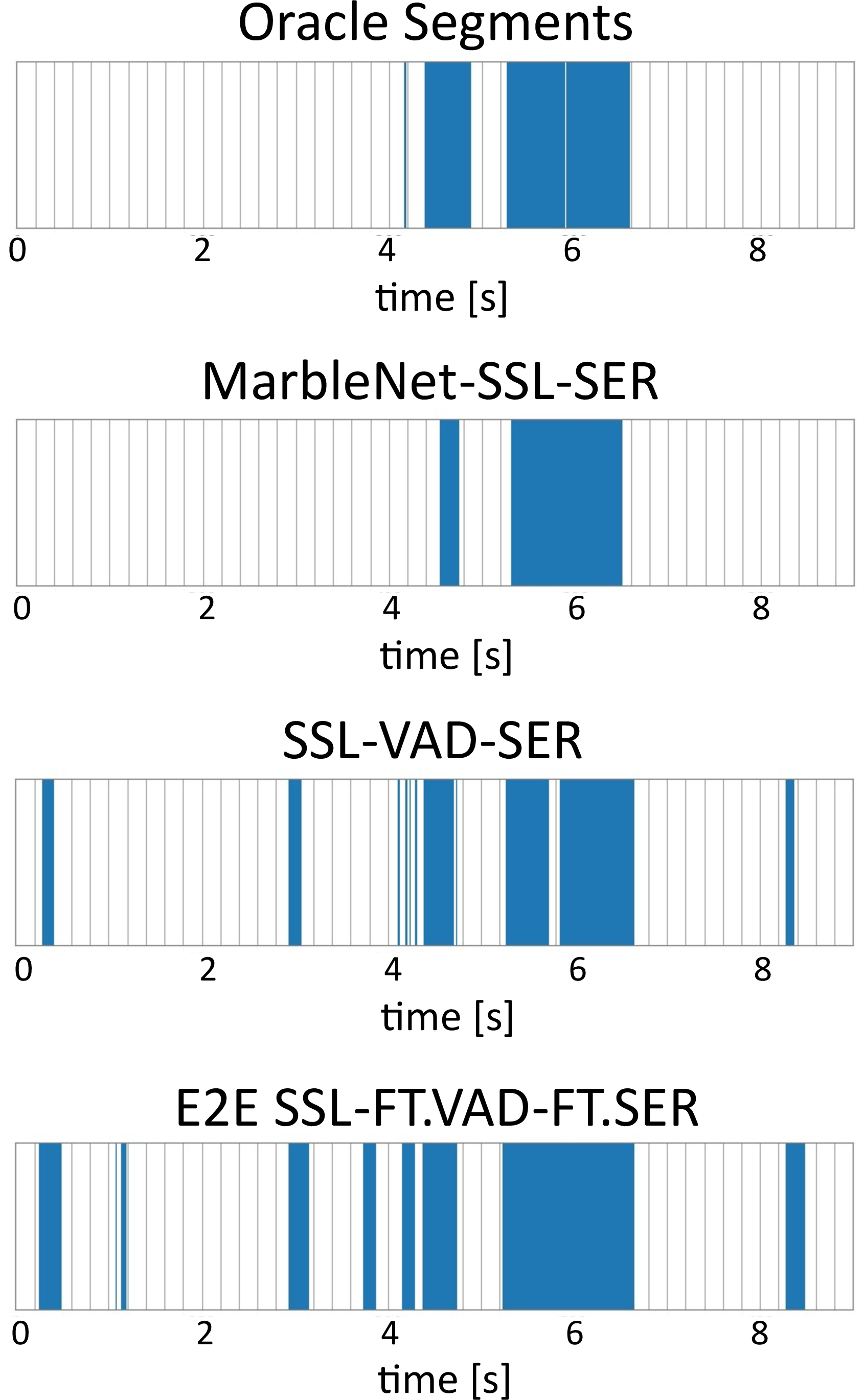}
        \subcaption{WavLM+}
        \label{fig:vad_wavlm}
    \end{minipage}
    \caption{Visualizations of the oracle segments annotated in the IEMOCAP and the VAD outputs in Conditions 3, 4, and 7 on HuBERT and WavLM+, sampled from results with an SNR of \SI{0}{\dB}.}  
    \label{fig:visualization}
\end{figure}

To investigate the impact of VAD outputs on SER performance, we analyzed both VAD performance and its outputs. 
Table~\ref{tab:vad} shows accuracy, precision and recall scores for VAD, and Fig.~\ref{fig:visualization} offers visualized examples of the VAD outputs, using HuBERT and WavLM+ under Conditions 3, 4, and 7.
Accuracy is the ratio of correctly predicted speech and non-speech segments to all segments. Precision is the ratio of correctly identified speech segments to all segments predicted as speech.
Recall is the ratio of correctly identified speech segments to all actual speech segments.

From the results, although MarbleNet exhibited higher precision, it often missed speech segments as depicted in Fig.~\ref{fig:vad_hubert}, which can lead to degrade SER performance.
In Condition 4 and 7, we cannot see a significant difference in the VAD performance as shown in Table~\ref{tab:vad}.
However, from the analysis of visualizations, it was observed that the fine-tuned VAD modules sometimes focused on important emotional parts as shown in Fig.~\ref{fig:vad_hubert}, or broadly detected segments to include more emotion despite containing more noise as shown in Fig.~\ref{fig:vad_wavlm}, depending on the input utterances and the SSL models.
These findings suggest that optimizing the VAD module for SER in an end-to-end manner does not necessarily lead to improvement in VAD performance.
Additionally, we observed that the SSL-VAD models tended to produce fragmented outputs as shown in Fig.~\ref{fig:vad_wavlm}.
This indicates that fine-tuning the SER module, which was pre-trained only on the original unsegmented utterances, with such flawed outputs likely contributed to improving SER performance in Conditions 6 and 7.

 
\subsection{Analysis of weights of the Featurizer}

\begin{figure}[t]
    \centering
    \begin{minipage}[b]{1\columnwidth}
        \centering
        \includegraphics[width=0.75\columnwidth]{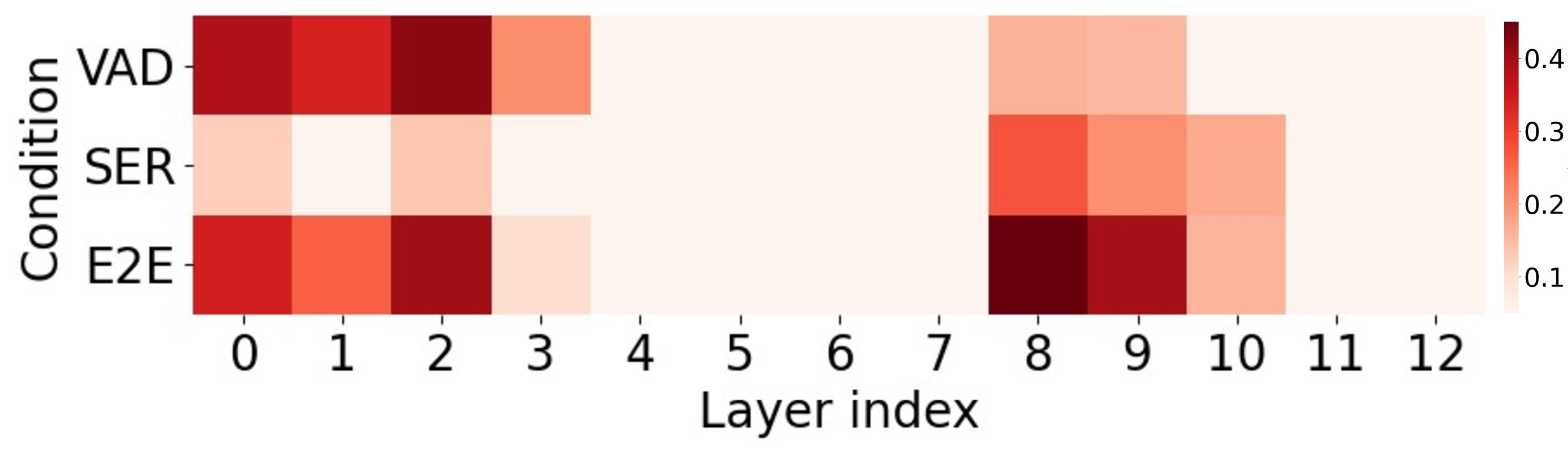}
        \subcaption{HuBERT}
        \vspace{5pt}
        \label{fig:a}
    \end{minipage}
    \begin{minipage}[b]{1\columnwidth}
        \centering
        \includegraphics[width=0.75\columnwidth]{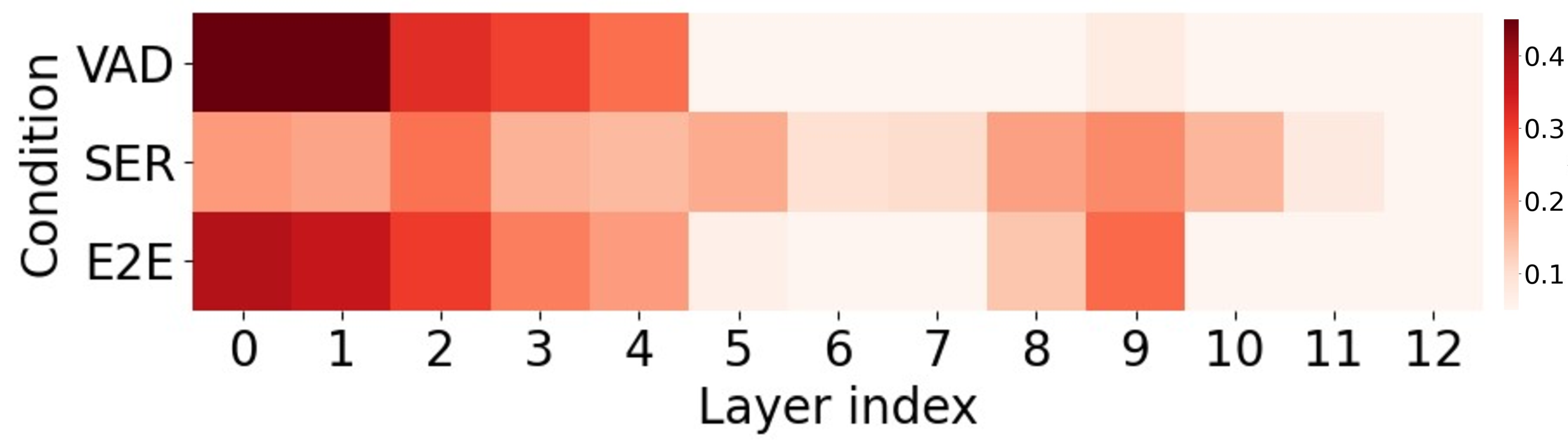}
        \subcaption{WavLM+}
        \label{fig:b}
    \end{minipage}
    \caption{Visualization of the Featurizer weights on HuBERT and WavLM+. The y-axis represents the weights of each Featurizer for SSL-VAD-SER (top), SSL-SER (middle), and E2E SSL-FT.VAD.FT-SER (bottom), while the x-axis represents different layers. Layer 0 corresponds to the input of the first Transformer layer.}  
    \label{fig:featurizer}
\end{figure}

We investigated the weights of the Featurizer in Conditions 4, 2, and 7, which were trained for VAD, SER, and both VAD and SER, respectively.
Fig.~\ref{fig:featurizer} shows the weights of different layers of the Featurizer on HuBERT and WavLM+. 
The results indicate that Layers 0-4 are useful for VAD, while Layers 8-10 are more effective for SER. 
These patterns are consistent with previous findings \cite{superb,wavlm,multitask}. 
We observe that our end-to-end method successfully emphasizes the features of Layers 8 and 9 for SER, as well as of Layers 0-4 for VAD, enabling the VAD module to take emotional features into account.


\section{Conclusions}

In this study, we presented a method that integrates VAD and SER modules using SSL features in an end-to-end manner. 
Our approach allows the VAD module to capture effective speech segments for SER, while making the SER module robust against flawed segments from the VAD module. 
Experimental results on the IEMOCAP dataset showed that our proposed method significantly improved SER performance in noisy environments.
Future work will include investigating the generalization ability of our proposed approach.

\clearpage

\bibliographystyle{IEEEtran}
\footnotesize
\bibliography{mybib}

\begin{thebibliography}{10}
\providecommand{\url}[1]{#1}
\csname url@samestyle\endcsname
\providecommand{\newblock}{\relax}
\providecommand{\bibinfo}[2]{#2}
\providecommand{\BIBentrySTDinterwordspacing}{\spaceskip=0pt\relax}
\providecommand{\BIBentryALTinterwordstretchfactor}{4}
\providecommand{\BIBentryALTinterwordspacing}{\spaceskip=\fontdimen2\font plus
\BIBentryALTinterwordstretchfactor\fontdimen3\font minus \fontdimen4\font\relax}
\providecommand{\BIBforeignlanguage}[2]{{%
\expandafter\ifx\csname l@#1\endcsname\relax
\typeout{** WARNING: IEEEtran.bst: No hyphenation pattern has been}%
\typeout{** loaded for the language `#1'. Using the pattern for}%
\typeout{** the default language instead.}%
\else
\language=\csname l@#1\endcsname
\fi
#2}}
\providecommand{\BIBdecl}{\relax}
\BIBdecl

\bibitem{review}
V.~Dimitrios and K.~Constantine, ``Emotional speech recognition: Resources, features, and methods,'' \emph{Computer Speech \& Language}, vol.~48, no.~9, pp. 1162--1181, 2006.

\bibitem{affective_computing}
C.-H. Wu, Y.-M. Huang, and J.-P. Hwang, ``Review of affective computing in education/learning: Trends and challenges,'' \emph{British Journal of Educational Technology}, vol.~47, no.~6, pp. 1304--1323, 2016.

\bibitem{HI}
R.~Cowie, E.~Douglas-Cowie, N.~Tsapatsoulis, G.~Votsis, S.~Kollias, W.~Fellenz, and J.~G. Taylor, ``Emotion recognition in human-computer interaction,'' \emph{IEEE Signal processing magazine}, vol.~18, no.~1, pp. 32--80, 2001.

\bibitem{healthcare}
N.~A. Vaidyam, H.~Wisniewski, J.~D. Halamka, M.~S. Kashavan, and J.~B. Torous, ``Chatbots and conversational agents in mental health: a review of the psychiatric landscape,'' \emph{The Canadian Journal of Psychiatry}, vol.~64, no.~7, pp. 456--464, 2019.

\bibitem{customer_service}
V.~Petrushin, ``Emotion in speech: Recognition and application to call centers,'' in \emph{Proc. ANNIE}, vol. 710, 1999, p.~22.

\bibitem{marketing}
R.~P. Bagozzi, M.~Gopinath, and P.~U. Nyer, ``The role of emotions in marketing,'' \emph{Journal of the academy of marketing science}, vol.~27, no.~2, pp. 184--206, 1999.

\bibitem{ser_conventional_1}
T.~L. Nwe, S.~W. Foo, and L.~C. De~Silva, ``Speech emotion recognition using hidden markov models,'' \emph{Speech communication}, vol.~41, no.~4, pp. 603--623, 2003.

\bibitem{ser_conventional_2}
A.~Milton, S.~S. Roy, and S.~T. Selvi, ``Svm scheme for speech emotion recognition using mfcc feature,'' \emph{International Journal of Computer Applications}, vol.~69, no.~9, pp. 34--39, 2013.

\bibitem{ser_conventional_3}
W.~Lim, D.~Jang, and T.~Lee, ``Speech emotion recognition using convolutional and recurrent neural networks,'' in \emph{Proc. APSIPA}, 2016, pp. 1--4.

\bibitem{ser_conventional_4}
Y.~Xie, R.~Liang, Z.~Liang, C.~Huang, C.~Zou, and B.~Schuller, ``Speech emotion classification using attention-based lstm,'' \emph{IEEE/ACM Transactions on Audio, Speech, and Language Processing}, vol.~27, no.~11, pp. 1675--1685, 2019.

\bibitem{transformer}
A.~Vaswani, N.~Shazeer, N.~Parmar, J.~Uszkorei, L.~Jones, A.~N. Gomez, L.~Kaiser, and I.~Polosukhin, ``Attention is all you need,'' in \emph{Proc. APSIPA}, 2016, pp. 1--4.

\bibitem{ssl_based_1}
L.~Pepino, P.~Riera, and L.~Ferrer, ``Emotion recognition from speech using wav2vec 2.0 embeddings,'' in \emph{Proc. IEEE Spoken Language Technology Workshop}, 2021, pp. 3400--3404.

\bibitem{ssl_based_2}
M.~Macary, M.~Tahon, Y.~Estève, and A.~Rousseau, ``On the use of self-supervised pre-trained acoustic and linguistic features for continuous speech emotion recognition,'' in \emph{Proc. IEEE Spoken Language Technology Workshop}, 2020, pp. 373--380.

\bibitem{ssl_based_3}
H.~Zou, Y.~Si, C.~Chen, D.~Rajan, and E.~S. Chng, ``Speech emotion recognition with co-attention based multi-level acoustic information,'' in \emph{Proc. ICASSP}, 2022, pp. 7367--7371.

\bibitem{wav2vec2}
A.~Baevski, H.~Zhou, A.~Mohamed, and M.~Auli, ``{wav2vec 2.0}: A framework for selfsupervised learning of speech representations,'' in \emph{Proc. NeurIPS}, vol.~33, 2020, pp. 12\,449--12\,460.

\bibitem{hubert}
W.-N. Hsu, B.~Bolte, Y.-H.~H. Tsai, K.~Lakhotia, R.~Salakhutdinov, and A.~Mohamed, ``{HuBERT}: Self-supervised speech representation learning by masked prediction of hidden units,'' \emph{IEEE/ACM TASLP}, vol.~29, pp. 3451--3460, 2021.

\bibitem{wavlm}
S.~Chen, C.~Wang, Z.~Chen, Y.~Wu, S.~Liu, Z.~Chen, J.~Li, N.~Kanda, T.~Yoshioka, X.~Xiao \emph{et~al.}, ``{WavLM}: Large-scale self-supervised pre-training for full stack speech processing,'' \emph{IEEE Journal of Selected Topics in Signal Processing}, vol.~16, no.~6, pp. 1505--1518, 2022.

\bibitem{finetuning_1}
Y.~Wang, A.~Boumadane, and A.~Heba, ``A fine-tuned wav2vec 2.0/hubert benchmark for speech emotion recognition, speaker verification and spoken language understanding,'' 2021.

\bibitem{finetuning_2}
L.-W. Chen and A.~Rudnicky, ``Exploring wav2vec 2.0 fine tuning for improved speech emotion recognition,'' in \emph{Proc. ICASSP}, 2023, pp. 1--5.

\bibitem{finetuning_3}
W.~Chen, X.~Xing, P.~Chen, and X.~Xu, ``Vesper: A compact and effective pretrained model for speech emotion recognition,'' \emph{IEEE Transactions on Affective Computing}, 2024.

\bibitem{superb}
S.-w. Yang, P.-H. Chi, Y.-S. Chuang, C.-I.~J. Lai, K.~Lakhotia, Y.~Y. Lin, A.~T. Liu, J.~Shi, X.~Chang, G.-T. Lin \emph{et~al.}, ``{SUPERB}: Speech processing universal performance benchmark,'' in \emph{Proc. NeurIPS}, vol.~33, 2020, pp. 12\,449--12\,460.

\bibitem{vad_review}
M.~Sharma, S.~Joshi, T.~Chatterjee, and R.~Hamid, ``A comprehensive empirical review of modern voice activity detection approaches for movies and {TV} shows,'' \emph{Neurocomputing}, vol. 494, pp. 116--131, 2022.

\bibitem{marblenet}
F.~Jia, S.~Majumdar, and B.~Ginsburg, ``Marblenet: Deep 1d time-channel separable convolutional neural network for voice activity detection,'' in \emph{Proc. ICASSP}, 2021, pp. 6818--6822.

\bibitem{vad_deeplearning_1}
N.~Wilkinson and T.~Niesler, ``A hybrid cnn-bilstm voice activity detector,'' in \emph{Proc. ICASSP}, 2021, pp. 6803--6807.

\bibitem{vad_deeplearning_2}
Q.~Yang, Q.~Liu, N.~Li, M.~Ge, Z.~Song, and H.~Li, ``Svad: A robust, low-power, and light-weight voice activity detection with spiking neural networks,'' in \emph{Proc. ICASSP}, 2024, pp. 221--225.

\bibitem{vad_conventional_1}
J.~Sohn, N.~S. Kim, and W.~Sung, ``A statistical model-based voice activity detection,'' \emph{IEEE signal processing letters}, vol.~6, no.~1, pp. 1--3, 1999.

\bibitem{vad_conventional_2}
J.-H. Chang, N.~S. Kim, and S.~K. Mitra, ``Voice activity detection based on multiple statistical models,'' \emph{IEEE Transactions on Signal Processing}, vol.~54, no.~6, pp. 1965--1976, 2006.

\bibitem{webrtcvad}
\BIBentryALTinterwordspacing
``{WebRTC VAD}.'' [Online]. Available: \url{https://webrtc.org/}
\BIBentrySTDinterwordspacing

\bibitem{ser_noise}
Y.~Huang, J.~Xiao, K.~Tian, A.~Wu, and G.~Zhang, ``Research on robustness of emotion recognition under environmental noise conditions,'' \emph{IEEE Access}, vol.~7, pp. 142\,009--142\,021, 2019.

\bibitem{ssl_layer}
A.~Pasad, J.-C. Chou, and K.~Livescu, ``Layer-wise analysis of a self-supervised speech representation model,'' in \emph{Proc. ASRU}, 2021, pp. 914--921.

\bibitem{ssl_vad}
B.~Karan, J.~J. van V{\"u}ren, F.~de~Wet, and T.~Niesler, ``A transformer-based voice activity detector,'' in \emph{Proc. Interspeech}, 2024, pp. 3819--3823.

\bibitem{silerovad}
\BIBentryALTinterwordspacing
“. V. p.-t. e.-g. v. a. d. V. n.~d. Silero~Team and language classifier, 2021. [Online]. Available: \url{https://github.com/snakers4/silero-vad}
\BIBentrySTDinterwordspacing

\bibitem{self_attention_pooling}
P.~Safari, M.~India, and J.~Hernando, ``Self-attention encoding and pooling for speaker recognition,'' in \emph{Proc. Interspeech}, 2020, pp. 941--945.

\bibitem{vad_ser}
M.~F. Alghifari, T.~S. Gunawan, M.~A. b.~W. Nordin, S.~A.~A. Qadri, M.~Kartiwi, and Z.~Janin, ``On the use of voice activity detection in speech emotion recognition,'' \emph{Bulletin of Electrical Engineering and Informatics}, vol.~8, no.~4, pp. 3607--3611, 2019.

\bibitem{multitask}
W.~u, C.~Zhang, and P.~C. Woodland, ``Integrating emotion recognition with speech recognition and speaker diarisation for conversations,'' in \emph{Proc. Interspeech}, 2023, pp. 941--945.

\bibitem{iemocap}
C.~Busso, M.~Bulut, C.-C. Lee, A.~Kazemzadeh, E.~Mower, S.~Kim, J.~N. Chang, S.~Lee, and S.~S. Narayanan, ``{IEMOCAP}: Interactive emotional dyadic motion capture database,'' \emph{Language resources and evaluation}, vol.~42, pp. 335--359, 2008.

\bibitem{musan}
D.~Snyder, G.~Chen, and D.~Povey, ``{MUSAN}: A music, speech, and noise corpus,'' arXiv:1510.08484, 2015.

\bibitem{librispeech}
V.~Panayotov, G.~Chen, D.~Povey, and S.~Khudanpur, ``{LibriSpeech}: An {ASR} corpus based on public domain audio books,'' in \emph{Proc. ICASSP}, 2015, pp. 5206--5210.

\bibitem{libri_light}
J.~Kahn, M.~Riviere, W.~Zheng, E.~Kharitonov, Q.~Xu, P.-E. Mazar{\'e}, J.~Karadayi, V.~Liptchinsky, R.~Collobert, C.~Fuegen \emph{et~al.}, ``{Libri-Light}: A benchmark for {ASR} with limited or no supervision,'' in \emph{Proc. ICASSP}, 2020, pp. 7669--7673.

\bibitem{gigaspeech}
G.~Chen, S.~Chai, G.~Wang, J.~Du, W.-Q. Zhang, C.~Weng, D.~Su, D.~Povey, J.~Trmal, J.~Zhang \emph{et~al.}, ``{GigaSpeech}: An evolving, multi-domain {ASR} corpus with 10,000 hours of transcribed audio,'' in \emph{Proc. INTERSPEECH}, 2021, pp. 3670--3674.

\bibitem{voxpopuli}
C.~Wang, M.~Riviere, A.~Lee, A.~Wu, C.~Talnikar, D.~Haziza, M.~Williamson, J.~Pino, and E.~Dupoux, ``{VoxPopuli}: A large-scale multilingual speech corpus for representation learning, semi-supervised learning and interpretation,'' in \emph{Proc. ACL-IJCNLP}, 2021, pp. 993–--1003.

\bibitem{adam}
D.~P. Kingma and J.~Ba, ``Adam: A method for stochastic optimization,'' in \emph{Proc. ICLR}, 2015.

\end{thebibliography}

\end{document}